\documentclass[12pt,preprint]{aastex}

\shorttitle{Seismological determination of flow speeds in coronal loops}
\shortauthors{Terradas et al.}

\begin{document}



\title{Seismology of transversely oscillating coronal loops with siphon flows}


\author{J. Terradas, I. Arregui}
\affil{Departament de F\'\i sica, Universitat de les Illes Balears, E-07122 Palma de Mallorca, Spain}
\email{jaume.terradas@uib.es}
\author{G. Verth, M. Goossens}
\affil{Centrum voor Plasma Astrofysica and Leuven Mathematical Modeling and
Computational Science Centre, KU Leuven, Celestijnenlaan 200B, 3001 Heverlee, Belgium}

\begin{abstract} There are ubiquitous flows observed in the solar atmosphere of
sub-Alfv\'{e}nic speeds, however after flaring and coronal mass ejection events
flows can become Alfv\'{e}nic. In this Letter, we derive an expression for the
standing kink mode frequency due to siphon flow in coronal loops, valid for both low
and high speed regimes. It is found that siphon flow introduces a linear spatially
dependent phase shift along coronal loops and asymmetric eigenfunctions. We
demonstrate how this theory can be used to determine the kink and flow speed of
oscillating coronal loops with reference to an observational case study. It is shown
that the presence of siphon flow can cause the underestimation of magnetic field
strength in coronal loops using the traditional seismological methods.
\end{abstract}


\maketitle

\section{Introduction}

The measurement of plasma flow speed in the solar atmosphere has traditionally
been estimated with Doppler shift using spectrometers onboard, e.g., SOHO and
Hinode \citep[e.g.,][]{brekkeetal97,winetal02}. Some attempts have also been
made by tracking features thought to be associated with flow using imagers
\citep{winetal01,chaeetal08}. It has been found that sub-Alfv\'{e}nic flows
around of 100 km s$^{-1}$ are ubiquitous in the corona. However, less frequent
but faster flows have been detected in the vicinity of flaring events and
coronal mass ejections \citep[e.g.,][]{innesetal01,harraetal05}, even into the
Alfv\'{e}nic regime of $10^3$ km s$^{-1}$ \citep[e.g.,][]{innesetal03}. Some of
the reported flows are of siphon type, i.e., the flow is unidirectional from one
end of the loop to the other. Spectroscopic identifications of siphon flows can
be found, for example, in \citet{teriacaetal04} with SUMER/SOHO and
\citet{tianetal08} with SUMER/SOHO and EIS/HINODE. Very clear identifications of
siphon flows through imaging observations were reported by \citet{doyleetal06}
with TRACE and \citet{tianetal09} with STEREO. 

As well as generating fast flows, flares and CMEs can also cause standing kink
oscillations in coronal loops \citep[see e.g.,][]{aschetal99,nakaetal99}, so it
is natural to develop magnetohydrodynamic (MHD) theory that models the interaction of these waves with
flow. Since the dynamics of coronal plasma are dominated by magnetic fields, in
general, flow direction will be magnetic field aligned. Using Doppler shift,
there may be large uncertainty in estimating flow speeds along coronal
structures such as loops due to the line of sight effects. In the most extreme case, if
the direction of flow is perpendicular to the line of sight, there will be no
flow detected at all. Furthermore, if a loop with field aligned flow is
oscillating with a kink mode this adds further difficulty in measuring the flow
speed using Doppler shift. In this Letter we want to demonstrate how the problem
of estimating the flow speed in oscillating coronal loops can be addressed by
implementing the technique of magnetoseismology.

Coronal seismology is an indirect way to obtain the magnitudes of fundamental
plasma parameters exploiting the observed oscillatory properties of the solar
atmosphere. This idea was proposed by \citet{uchida1970} and \citet{robetal84}
and up to now, has successfully provided estimates of the magnetic field
strength \citep{nakaof01,vandetal08a} and density scale height
\citep{andriesetal05a,verthetal08} in the corona. Also,
\citet[][]{arreguietal07,goossetal08} established lower limits to the
sub-resolution transverse inhomogeneity in coronal loops, and upper limits to
internal Alfv\'en speeds.

In this Letter, for the first time, we demonstrate how the flow speed along
oscillating coronal loops can be determined from observations of standing kink
modes. We theoretically investigate the effect of a constant siphon flow on
stationary transverse waves and link this model to an observational case study.

\section{Equilibrium model}\label{nonequil}

Consider an equilibrium model of a cylindrical axis-symmetric flux tube of
radius $R$ with constant axial magnetic field $B_0$, and a density contrast of
$\rho_{\mathrm i}/\rho_{\mathrm e}$. The subindex ``${\mathrm i}$'' and
``${\mathrm e}$'' refer to the internal and external part of the tube,
respectively. The length of the tube is $L$ and it is assumed that inside there
is a unidirectional constant flow $U$ and no flow outside.

The stationary solution is constructed by superposing two
propagating waves traveling in opposite directions. The effect of flow on propagating waves causes a frequency
Doppler shift, a phenomena studied in the past by many authors, e.g.,
\citet{goossensetal92}. In particular it was found that, in the zero-$\beta$
limit, a valid approximation for the coronal plasma, frequency is given
by the following expression,
\begin{eqnarray}\label{freq} \omega= k \frac{\rho_{\mathrm i}}{\rho_{\mathrm i} +
\rho_{\mathrm e}} U \pm k\, c_{\mathrm{k}} \sqrt{
1-\frac{\rho_{\mathrm i}\rho_{\mathrm e}}{(\rho_{\mathrm i}+\rho_{\mathrm
e})^2}\frac{U^2}{c_{\mathrm{k}}^2}}, \end{eqnarray}
\noindent
where $c_{\mathrm{k}}$ the kink speed for a static equilibrium within the thin tube regime ($R\ll L$). The corresponding
eigenfunctions \citep[see for example][]{edrob83} are expressed in terms of
Bessel functions in the radial direction and have a sinusoidal dependence in the
longitudinal direction.

To find a stationary solution, we impose that the forward ($+$ sign) and backward
($-$ sign) waves must have the same frequency. This condition can be only satisfied if the
corresponding longitudinal wavenumbers are different, and according to
Eq.~(\ref{freq}) are given by,
\begin{eqnarray} k_\pm=
\frac{\omega}{\frac{\rho_{\mathrm i}}{\rho_{\mathrm i}+\rho_{\mathrm e}} U\pm
c_{\mathrm{k}}\sqrt{
1-\frac{\rho_{\mathrm i}\rho_{\mathrm e}}{(\rho_{\mathrm i}+\rho_{\mathrm
e})^2}\frac{U^2}{c_{\mathrm{k}}^2}}}.\end{eqnarray}
Note that in general $k_+>0$, and $k_-<0$ (we do not consider KH-unstable modes
here, i.e., we assume the square root is always positive).

We concentrate on the $z-$dependence of the stationary solution, neglecting
the radial dependence because we are assuming
the thin tube limit. The study of the full MHD eigenvalue problem
would require a more involved analysis. The transversal displacement of
the tube axis is thus
\begin{eqnarray}\label{eigenfunct}
\xi(t,z)=A\sin\left(\omega t
-k_+ z\right)+ B\sin\left(\omega t -k_- z\right),
\end{eqnarray}
with $A$ and $B$ constants determined by boundary conditions. Imposing line-tying conditions at the footpoints, i.e.,
\begin{eqnarray} \xi(t,z=0)=0,\\ \xi(t,z=L)=0,
\end{eqnarray}
from the first condition we find that $A=-B$. Using this relation
and the second condition, we obtain the following dispersion relation
\begin{eqnarray}\label{omegaeigen}
\omega=k_0\,c_{\mathrm{k}} \frac{\left(
1-\frac{\rho_{\mathrm i}}{\rho_{\mathrm i}+\rho_{\mathrm
e}}
 \frac{U^2}{c_{\mathrm{k}}^2}\right)}{\sqrt{
1-\frac{\rho_{\mathrm i} \rho_{\mathrm e}}{\left(\rho_{\mathrm i}+\rho_{\mathrm
e}\right)^2}
 \frac{U^2}{c_{\mathrm{k}}^2}}} ,
\end{eqnarray}
where  \begin{eqnarray}\label{k0} k_0=n\frac{\pi}{L}, \,\,\, n= 1, 2, \ldots.
\end{eqnarray} \noindent The natural eigenfrequencies of the flux tube are given
by Eq.~(\ref{omegaeigen}) \citep[see][for the equivalent result for a pure
Alfv\'en wave, and the recently published work of \citet{ruderman10}, for a more
general study]{taroyan09}. This expression gives the explicit
dependence of frequency on $U$. If we compare with the eigenfrequencies of the
static case, $ \omega=k_0 c_{\mathrm{k}}$, it turns out that flow always leads
to a frequency reduction, i.e., an increase in the period of oscillation.

We now turn our attention to the time-dependence of the amplitude of
oscillations along the loop. Using the fact that $A=-B$ and
Eq.~(\ref{omegaeigen}), we have that Eq.~(\ref{eigenfunct}) can be written, after
some algebra, in the following elementary form
\begin{eqnarray}\label{eigenfuncts}
\xi(t,z)=C  \sin k_0 z \cos \left(\omega t +k_U z\right),
\end{eqnarray}
where $C$ is an arbitrary constant, and we have introduced the wavenumber due to
flow,
\begin{eqnarray}\label{kU}
k_{U}=k_0\frac{\rho_{\mathrm i}}{\rho_{\mathrm
i}+\rho_{\mathrm e}}\frac{U}{c_{\mathrm{k}}}\frac{1}{\sqrt{
1-\frac{\rho_{\mathrm i} \rho_{\mathrm e}}{\left(\rho_{\mathrm i}+\rho_{\mathrm
e}\right)^2}
 \frac{U^2}{c_{\mathrm{k}}^2}}}.
\end{eqnarray}

\noindent When there is no flow, $U=0$, we recover the standing wave pattern of
the static case, being a solution separable in time and space. When there is
flow, the solution is no longer separable but the interpretation is straight
forward. The term with the cosine in Eq.~(\ref{eigenfuncts}) represents a
propagating wave traveling in the opposite direction of the flow with an
effective wavenumber, which is simply linearly proportional to flow speed in
the slow flow regime ($U \ll c_{\mathrm{k}}$). While the term with the sine is
just the envelope of the traveling wave, satisfying the line-tying conditions
at the footpoints. Key to the present investigation, Eqs.~(\ref{eigenfuncts})
and~(\ref{kU}) show there are two main observational signatures of siphon flow
in a coronal loop kink standing wave: \newline

Signature 1.-- There is a linear phase dependence of the standing kink mode along the loop.
\newline

Signature 2.-- In one full period the eigenfunctions of the standing kink mode
will mostly exhibit an asymmetry about the center of the loop. \newline

\begin{figure}\centering
\includegraphics[width=8.cm]{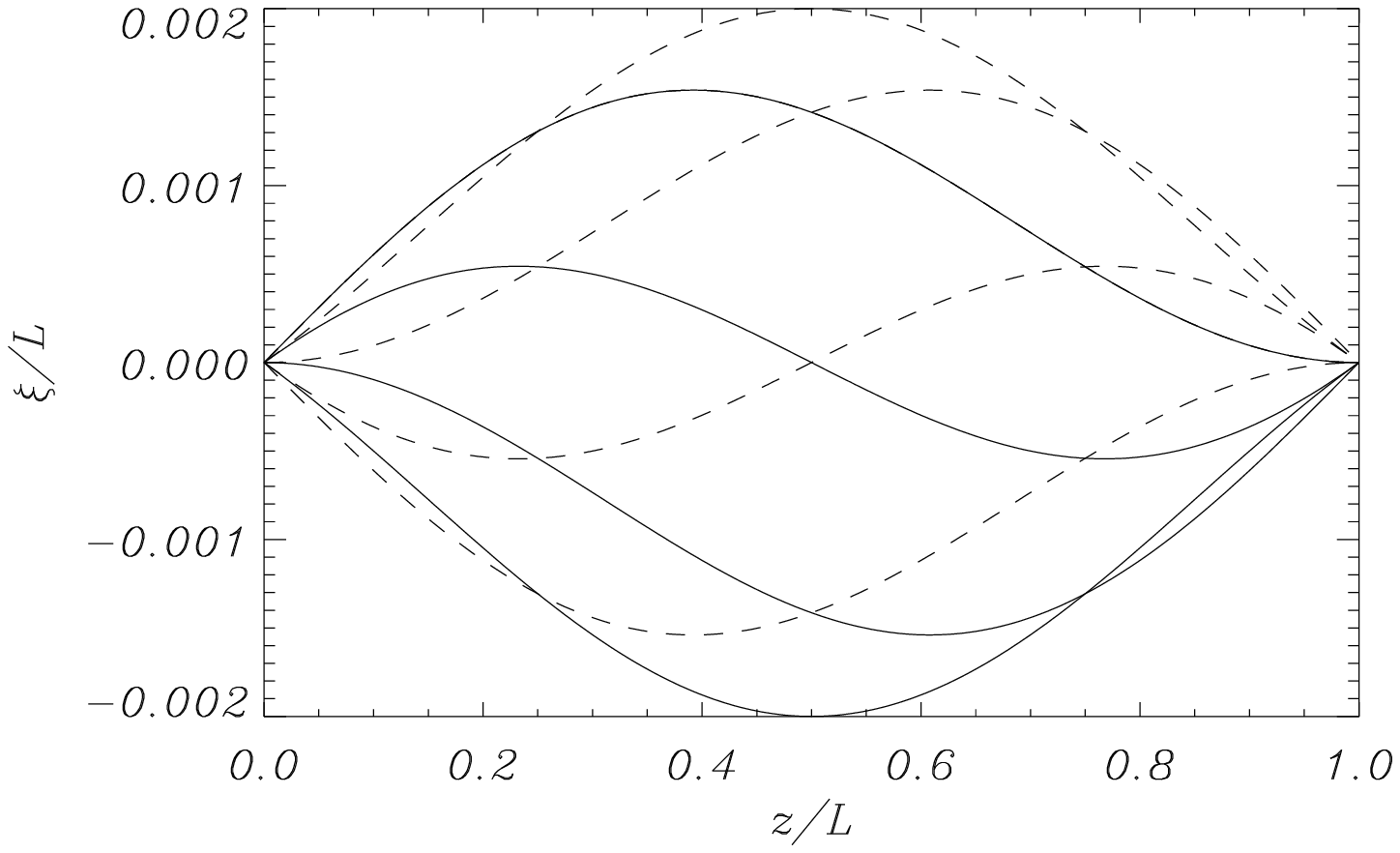}
\caption{\small Time evolution of the fundamental mode ($n=1$) as a function of
position along the loop for the case with $U=0.5
c_{\mathrm{k}}$. The curves represent the eigenfunction at different times (the interval is $P/8$,
where $P=2\pi/\omega$). The continuous lines represent
motion of the loop in the negative direction, while dashed lines
represent motion in the positive direction.}  \label{eigenplot}
\end{figure}

\noindent If either Signature 1 or 2 are present in data, then this may be an
indication there is a siphon flow present but if both are present this is a
stronger indication of siphon flow. Figure~\ref{eigenplot} illustrates the
dependence of amplitude along a coronal loop at different times due to constant
siphon flow, illustrating Signature 2 of siphon flow. It is worth noting that in
the absence of siphon flow Signature 1 may simply indicate the presence of a
propagating kink wave. The correct interpretation of data must be done on a case
by case basis, taking into account loop geometry and all estimated wave
parameters.

\section{Expressions for siphon flow and kink speed}
\label{U and Ck}
Usually the data analysis of standing kink oscillations is based on a fit of the
loop displacement at a given point along the loop of the following form
\begin{eqnarray}\label{fitobs} \xi=\xi_0 \cos \left(\omega t +\phi\right)
e^{-t/\tau_\mathrm {D}}.
\end{eqnarray}

\noindent The parameters that are determined from the fit are $\xi_0$, $\omega$,
$\phi$, and $\tau_\mathrm {D}$. The damping time, $\tau_\mathrm {D}$,  is not of
interest in this work since a damping mechanism is most likely required to produce
the attenuation of the signal with time. Regarding the other parameters it
is important to point out that it is possible, depending on the data set, to derive their values
along some portion of the loop and not just at a single point \citep[see e.g.,][]{verwetal10}.

Usually the fundamental mode is reported in the observations, meaning that $n=1$
in Eq.~(\ref{k0}) but there is also possible evidence of higher overtones
\citep[see e.g.,][]{verwetal04}. The loop length, $L$, can be estimated using a
circular shape or a 3D reconstruction, meaning that $k_0$, is well determined
(up to the uncertainties in $L$).

When there is no flow, the determination of $\omega$ (calculated using the fit)
allows us, together with the value of $k_0$, to calculate the kink speed of the
tube, i.e., $c_{\mathrm{k}}=\omega/k_0$. However, with only estimates for
$\omega$ and $k_0$, we cannot determine the kink speed if flows are present. We
also need additional information about the phase $\phi$ and plasma densities,
$\rho_\mathrm{i}$ and $\rho_\mathrm{e}$. From the Signature 1 of siphon flow in
Section \ref{nonequil}, phase should show a linear dependence with position
along the loop with gradient $k_U$ and this can be estimated with a linear fit
to data. Also, the values of $\rho_\mathrm{i}$ and $\rho_\mathrm{e}$ can be
estimated from observed intensity, since emission measure is approximately
proportional to density squared if one considers the intensity of a clean
emission line formed in the transition region or corona. In imaging
observations, usually a passband is wide enough to include several emission
lines with different formation temperatures and the integrated intensity of
this passband is no longer simply proportional to the density squared. The
density diagnostics is possible if you have observations of density sensitive
line pairs. But this usually needs spectroscopic observations (e.g., SUMER,
CDS, EIS).

Once $\omega$, $k_0$, $k_U$, $\rho_\mathrm{i}$ and
$\rho_\mathrm{e}$ are determined from observation, it is straight forward to
calculate the two magnitudes of interest, namely $U$ and $c_{\mathrm{k}}$. We
only need to combine Eq.~(\ref{omegaeigen}) and Eq.~(\ref{kU}) to find
\begin{eqnarray}\label{Uobs} U=\frac{\rho_{\mathrm i}+\rho_{\mathrm
e}}{\rho_{\mathrm i}} \frac{k_U}{k_0^2-k_U^2}\,\omega, \end{eqnarray}
and
\begin{eqnarray}\label{ckobs} c_{\mathrm
k}^2=\frac{1}{2}\frac{\omega^2}{k_0^2}+\frac{\rho_{\mathrm i}}{\rho_{\mathrm
i}+\rho_{\mathrm
e}}U^2+\frac{1}{2}\sqrt{\frac{\omega^2}{k_0^2}\left(\frac{\omega^2}{k_0^2}+4
\frac{\rho_{\mathrm i}^2}{\left(\rho_{\mathrm
i}+\rho_{\mathrm
e}\right)^2}U^2\right)}.\nonumber\\
\end{eqnarray}
Note that Eq.~(\ref{Uobs}) contains the information about the flow
direction as well, if $k_U>0$ ($k_U<0$) then we have that $U>0$ ($U<0$) assuming
that $k_0>k_U$.

\section{Observational case study}\label{obs}

There have been a number of observational studies which have shown a gradient
in phase along coronal loops oscillating with the standing kink mode, e.g.,
\citet{verwetal04}, \citet{vandetal07} and \citet{demoortbrady07}. However, it is
unclear from these studies whether the observed phase difference is due to
siphon flow or wave propagation since there is only evidence of Signature 1 in
these cases. For this reason we focus instead on the particular case study of
\citet{verwetal10}, since both Signature 1 and 2 are present in the data,
providing a stronger argument for the siphon flow interpretation. Using combined
TRACE and EIT observations, \citet{verwetal10} analyzed the standing kink mode
generated in a coronal loop that was part of a large arcade. The coronal loop,
positioned off-limb, starts oscillating after an M1.5 GOES level flare and
associated CME occurs at about 11:04 UT. Regarding Signature 1,
\citet{verwetal10} fitted a linear function to the observed phase along a 120 Mm
portion of the loop (about 18\% of the total loop length total $L=680$ Mm) and
this is shown in Figure 6 of their paper. Evidence of Signature 2, an asymmetry
in the amplitude about the loop half length is shown in their Figure 8.

\citet{verwetal10} perform a fit to the loop displacement as in Eq.~(\ref{fitobs}). The values of the
parameters of interest (see Table I of their paper) are $\omega=2.60\times 10^{-3}$
s$^{-1}$ ($P=2418$ s),
$k_0=4.62\times 10^{-3}$ Mm ($L=680$ Mm, and $n=1$). The authors also provide a linear fit to the phase as a
function of position along the loop (denoted by $s$) which is
\begin{eqnarray}\label{fitphase1}
\phi(s)=0.93-3.67\times 10^{-3}\,\left(\frac{s}{1 \mathrm{\, Mm}}- 80\right ) \,\,\, [\mathrm {rad}],
\end{eqnarray}
meaning that $k_U=-3.67\times 10^{-3}$ Mm$^{-1}$. Using Eq.~(\ref{Uobs}) we
find that $U=-1445$ km s$^{-1}$, while from Eq.~(\ref{ckobs}) we obtain that
$c_{\mathrm{k}}=1610$ km s$^{-1}$ (assuming a density contrast of
$\rho_{\mathrm i}/\rho_{\mathrm e}=5$). Using the static model we find that
$c_{\mathrm{k}}=563$ km s$^{-1}$, almost three times smaller than the inferred
value assuming a siphon flow. The value of the flow speed is negative, meaning
that the flow would be towards the footpoint shown in the lower image of Figure 1 of \citet{verwetal10}.

It is relatively simple to estimate the errors in the calculations of the flow
and  kink speeds. These errors are due to the uncertainties in the quantities
that appear in Eqs.~(\ref{Uobs}) and (\ref{ckobs}). Typically the errors are
$\delta L=0.05L$, $\delta P=0.05 P$, $\delta (\rho_{\mathrm e}/\rho_{\mathrm
i})=0.4$ and also assume that $\delta k_U=0.10 k_U$. Applying the error
propagation formula in Eqs.~(\ref{Uobs}) and (\ref{ckobs}) we find that the
flow is $1445 \pm 755$ km s$^{-1}$, and the kink speed is $1610\,\pm 903$ km
s$^{-1}$. In this example, the errors in the speeds are not small, and are
specially sensitive to the uncertainty in $k_U$. It is worth noting that the
calculated speeds and errors are based on the assumption  of constant flow
speed along the loop. It is clear that, for this particular example, the
assumption of a constant siphon flow significantly increases the inferred kink
speed (almost by a factor of 3). Therefore the presence of siphon flow has an
important implication for the method of estimating magnetic field strength
along coronal loops by \citet{nakaof01}, which assumes a static equilibrium. In
the present example the magnetic field strength would therefore be
underestimated by a factor of 3.

Assuming there is a siphon flow present, then the flow is in the fast
(Alfv\'{e}nic) flow regime of $10^3$ km s$^{-1}$. The oscillation event analyzed
here takes place in a coronal loop arcade in the vicinity of an M class flare
and CME. In similar events, fast flow signatures have often been measured in
Doppler shift \citep[e.g.,][]{innesetal01,harraetal05}. After an X class flare
\citet{innesetal03} measured similar Alfv\'{e}nic flow speeds in the range
$800-1000$ km s$^{-1}$ in a coronal arcade using combined SUMER and TRACE
observations, and it must be emphasized these Doppler shifts are likely to be
underestimates of true field aligned flows speeds due to line of sight effects.

Alfv\'enic flow would cause significant asymmetry about the loop half length in the
eigenfunction (Signature 2) along the loop. Interestingly, the estimated
eigenfunction derived by \citet{verwetal10} using EIT, shows that the fundamental
mode has such an asymmetry. This is clearly seen in Figure 8(a) (see the region
$s/L\in [0,0.4]$) where \citet{verwetal10} fit the observed displacement (solid
line) with that predicted assuming a planar loop model (dashed line). However,
according to our model, this asymmetry can be explained by fast axial flow (see
Figure~\ref{eigenplot}).

\section{Summary and Conclusions}

Using a simple model of a coronal loop,  we have derived an analytical
expression for the frequency of oscillation of a thin tube with a constant
siphon flow, valid for both slow and fast flow regimes. It was found that
calculation of the kink speed and estimation of magnetic field assuming a static
model give values which are smaller than the ones obtained in the presence of
flow. This questions the validity of traditional seismological methods assuming
a static background to estimate the magnetic field along coronal loops.

Contrary to the static case, different positions along the tube oscillate with a
different phase, in agreement with the recent results of \citet{ruderman10}.
However, we have pointed out in our work the importance of the phase shift,
since it has a linear behavior with the axial coordinate along coronal loops and
in the low flow regime is simply linearly proportional to the flow speed.
Furthermore, the presence of flow breaks the symmetry of the eigenmodes about
the loop half length.

According to our model, assuming the linear phase shift reported along the
post-flare coronal loop analyzed by \citet{verwetal10} is caused by a siphon flow,
the flow would be in the Alfv\'{e}nic regime. This is not completely unreasonable
since, in the dynamic post-flare environment of coronal loops such fast flows have
been estimated using Doppler shift and feature tracking techniques
\citep[e.g.,][]{innesetal01,innesetal03,harraetal05}. In any case, other cases need
to be investigated in order to prove that this technique can be also used with
sub-Alfv\'enic flows.

The most important observed parameter that determines the value of the flow (and
kink speed) in our model is the value of $k_U$. The value of this parameter was
estimated using only a small portion of the loop. If we determine the phase
dependence along the whole loop (SDO should help in this regard) we would have a
much more accurate determination of the speeds and a significant reduction in
the errors. This more complete information will also allow us to test whether
the flow is unidirectional or not, according to the sign of $k_U$ at the
footpoints.

In reality flows along coronal loops are much more complex than the simple model
proposed in this Letter. In particular, they are probably non-symmetric respect
to the loop apex, and more important, they are usually transient \citep[see for
example,][]{doyleetal06,tianetal09}. Nevertheless, we must understand the
simplest possible configuration before additional complications are added. In
this regard, we have made the first step in assessing the effect of constant
siphon flows on standing kink oscillations.

\begin{acknowledgements}{}\small

J.T. acknowledges the Universitat de les Illes Balears for a postdoctoral position
and the funding provided under projects AYA2006-07637 (Spanish Ministerio de
Educaci\'on y Ciencia) and FEDER funds. G.V. and M.G. acknowledge support from
K.U.Leuven via GOA/2009-009 and financial support from UIB during their stay at this
university. Authors also acknowledge M. Ruderman and E. Verwichte for their helpful
comments and the anonymous referee for his/her constructive suggestions.

\end{acknowledgements}


\end{document}